\documentclass[aps,prb,twocolumn,superscriptaddress,raggedbottom,showpacs,floatfix]{revtex4}
\usepackage{amsmath,amsfonts,amssymb,amsthm}
\usepackage{color}
\pdfoutput=1
\usepackage{graphicx,latexsym}

\begin{document}

\title{Turnstile pumping through an open quantum wire}

\author{Cosmin Mihai Gainar}
\affiliation{Science Institute, University of Iceland, Dunhaga 3, IS-107 Reykjavik, Iceland}
\author{Valeriu Moldoveanu}
\affiliation{National Institute of Materials Physics, P.O. Box MG-7,
Bucharest-Magurele, Romania}
\author{Andrei Manolescu}
\affiliation{School of Science and Engineering, Reykjavik University, Menntavegi 1,
IS-101 Reykjavik, Iceland}
\author{Vidar Gudmundsson}
\affiliation{Science Institute, University of Iceland, Dunhaga 3, IS-107 Reykjavik, Iceland}
\affiliation{Physics Division, National Center for Theoretical Sciences,
             PO Box 2-131, Hsinchu 30013, Taiwan}

\begin{abstract}

      We use a non-Markovian generalized master equation (GME) to describe
      the time-dependent charge transfer through a parabolically confined
      quantum wire  of a finite length coupled to semi-infinite quasi
      two-dimensional leads.  The quantum wire and the leads are in
      a perpendicular external magnetic field.  The contacts to the
      left and right leads depend on time and are kept out of phase
      to model a quantum turnstile of finite size. The effects of the
      driving period of the turnstile, the external magnetic field, the
      character of the contacts, and the chemical potential bias on the
      effectiveness of the charge transfer of the turnstile are examined,
      both in the absence and in the presence of the magnetic field.
      The interplay between the strength of the coupling and the strength
      of the magnetic field is also discussed.  We observe how the edge
      states created in the presence of the magnetic field contribute
      to the pumped charge.

\end{abstract}


\maketitle

\section{Introduction}

The time-dependent properties of semiconductor nanostructures and
their response to electric pulses are currently being studied
through transient current measurements and in a pump-and-probe
configuration.\cite{Tarucha,Lai,Naser} Along with these experimental
developments theoretical schemes for the description of time-dependent
transport emerged.  The methods include the non-equilibrium Keldysh-Green
function formalism,\cite{Stefanucci,Moldo} scattering theory,
\cite{Gudmundsson} and more recently the generalized master equation
(GME) adapted for electronic transport.\cite{Harbola,NJP1,Welack1}

These methods were used mostly for studying the transient currents
generated by a time-dependent potential applied on the sample or by a
time-dependent coupling between the leads and the sample.  An example
in the first category is the time dependent pumping of electrons
through a small open system.\cite{Stefanucci}  In the second category
we mention the transient currents and the geometrical effects imposed
by the lateral confinement of the sample.\cite{NJP1, NJP2} 
In another recent work we investigated the modulation of the drain
current when a sequence of square pulses is applied to the source probe
connected to a quantum dot and a short quantum wire described within
lattice model.\cite{PRB1}  That study was motivated by the experiments
of Naser {\it et al}.\cite{Naser}

In the present work we further exploit the GME method and study
the transport properties of a quantum wire operating in a turnstile
regime. The turnstile pump is a single-electron device where the sample
is periodically connected and disconnected with the left and right lead
respectively, but with a relative phase shift.  It was experimentally
created by Kouwenhoven {\it et al.}\cite{TSP} by modulating in time the
two tunneling barriers between a quantum dot and two leads. The electrons
were driven by a finite bias between the leads.  This setup is different
from a quantum pump where a current is generated by asymmetric external
oscillations, but without a bias.  In the experiment of Kouwenhoven {\it
et al.} the barrier heights oscillate out of phase in the following
sense: On the first half-cycle electrons enter from the source probe
in the system but there is no current in the drain probe because the
corresponding tunneling barrier is high enough to prevent this. In the
second half-cycle the source is disconnected, the drain contact opens,
and a discharge of the dot follows. It was found that an integer number
of electrons are transmitted through the structure in each pumping cycle.

More recently, due to the general interest in applications of
nanoelectronic devices, more complex turnstile pumps have been studied
by numerical simulations, like one-dimensional arrays of junctions
\cite{Mizugaki} or two-dimensional multidot systems. \cite{Ikeda}
In the present paper we predict that the turnstile operation can also 
be performed in a quantum wire sample and in an external magnetic
field as well.  Our results are obtained using a parabolic lateral
confinement model both for the quantum wire and for the leads.  We put
a special effort on describing the lead-sample contacts.  In a previous
work we studied the turnstile transport through a sample described
by a lattice (tight-binding) model.  The sample was a one-dimensional
system with two or three sites and the transport calculations were done
using nonequilibrium Keldysh-Green functions.\cite{TSPPRB} The quantum
wire considered in this work is much more complex. We start from the
single-particle Hamiltonian of a two-dimensional wire of length $L_x$
parabolically confined along the $y$ direction and with hard-wall
conditions at $\pm L_x/2$. The eigenfunctions of the Hamiltonian were
described in detail in Ref.\ \onlinecite{Gudmundsson} and will not be
repeated here.

The material is organized as follows: In Section \ref{model} we briefly
review the main equations of the model and of the GME method, Section
\ref{calc} is devoted to the numerical results for zero magnetic field
(\ref{Beq0}), in the presence of a magnetic field (\ref{Bne0}), and to the
edge states (\ref{ES}). The conclusion are given in Section \ref{conc}.

\section{The model}
\label{model}
We consider an isolated finite quantum wire of length
$L_x=300$ nm, extended in the $x$-direction.  The width of the wire is
defined by a parabolic confinement potential in the $y$-direction with
the characteristic energy $\hbar\Omega_0 = 1.0$ meV.  The quantum wire is
terminated at $\pm L_x/2$ with hard wall potentials.  In addition, we
have two semi-infinite leads, one extended from  $-L_x/2$ to $-\infty$,
and the other one from $+L_x/2$ to $+\infty$.  Both have a parabolic
confinement in the $y$ direction with an energy of 0.8 meV, and are
also terminated at $\pm L_x/2$ with hard walls.  The leads and the finite
quantum wire, or the system, are all subjected to an external constant
magnetic field $\mathbf{B}=B\mathbf{\hat{z}}$.  The length of the wire,
$L_x$, and the magnetic length modified by the parabolic confinement
$a_w=\sqrt{\hbar/(m^*\Omega_w)}$, with $\Omega_w^2=\Omega_0^2+\omega_c^2$
and the cyclotron frequency $\omega_c=eB/(m^*c)$, are convenient length
scales in the calculations.  We assume the GaAs effective mass, $m^*=0.067m_e$.
The many-electron Hamiltonian of the system composed by the semi-infinite leads 
and the finite quantum wire, but isolated from each other, is:
\begin{equation}
      H(t)=\sum_a E_a d^\dagger_a d_a
      +\sum_{q,l=\mathrm{L,R}} \epsilon_l(q) c^\dagger_{ql} c_{ql},
\end{equation}
where an electron in the system is created (annihilated) by the operators 
$d^\dagger$ ($d$), and in the leads by $c^\dagger$ ($c$). $E_a$ are
the energies of the single-electron states labeled with $a=1,2,3,...$
in increasing order, $\epsilon_l(q)$ is the energy spectrum of the
left and right leads, labeled as $l=\mathrm{L}$ and $l=\mathrm{R}$
respectively, and $q$ representing a discrete label of subbands and a
continuous quantum number labeling states within each subband.  At $t=0$
the system is coupled to the leads with the Hamiltonian
\begin{equation}\label{H_tun}
      H_\mathrm{T}(t)=\sum_l\chi_l(t)\sum_{q,a}
      \left\{T^l_{qa}c_{ql}^\dagger d_a + (T^l_{qa})^*d_a^\dagger c_{ql} \right\},
\end{equation}
with $\chi_l(t)$ describing the time-dependence of the coupling, such that 
$\chi_l(t<0)=0$,  and $T^l_{qa}$ describing the coupling strength of state 
$a$ and $q$ in the system and the leads, respectively.

The coupling is defined by a nonlocal overlap integral of the two states in the region
of contact around $\pm L_x/2$.  The coupling coefficients are defined phenomenologically 
with the tensor \cite{NJP2}
\begin{equation}
      T^l_{aq} = \int_{\Omega_S^l\times \Omega_l} d{\bf r}d{\bf r}'
      \left(\Psi^l_q ({\bf r}') \right)^*\Psi^S_a({\bf r})
      g^l_{aq} ({\bf r},{\bf r'})+h.c.,
\label{T_aq}
\end{equation}
where a nonlocal-overlap of the wave functions in the system and the leads
is modeled by
\begin{align}
      g^l_{aq} ({\bf r},{\bf r'}) =
                   g_0^l&\exp{\left[-\delta_1^l(x-x')^2-\delta_2^l(y-y')^2\right]}\nonumber\\ 
                   &\exp{\left(\frac{-|E_a-\epsilon^l(q)|}{\Delta_E^l}\right)}.
\label{gl}
\end{align}
The strength of the coupling between the leads and the sample is defined
by the parameter $g_0^l$, which captures the tunneling rate at the
contact between each lead and the sample, and also by the parameters
$\delta_1^l$, $\delta_2^l$ and $\Delta_E^l$ which adjust the spatial overlap of
lead and sample wave functions in the contact region.  Since all we have
from our model are the wave functions derived for the uncoupled subsystems,
this intuitive ansatz, Eqs.\ (\ref{T_aq}) and (\ref{gl}), is a convenient
way of describing the coupling.  In our calculations these coupling parameters
will be the same for both leads and hence the label $l$ will be omitted.

The system can be subjected to a bias $\Delta\mu =\mu_L-\mu_R$
and in order to reduce the number of many-electron states (MES)
to a reasonable number in the following calculations we limit the number 
of single-electron states (SES)
for the particular calculation by selecting a window
of relevant states around the bias window, i.e.\ $[\mu_R-\Delta ,
\mu_L+\Delta ]$, such that the transport properties are not changed
significantly by extending the window.  We consider these states
relevant for the transport, or ``active'', while all the other states are
``frozen'', being either permanently occupied or permanently empty.
In addition to the electrons frozen in the states below the active window
(which are not included in the transport calculation), we also assume
the lowest active state occupied at the moment $t=0$ {\it i.\ e.}\ when the contacts
begin to operate.  In this way the transient phase is shorter than if
we would assume an empty active window, and we can spend less computing
time until the system reaches the periodic state.

The time-evolution of the total system - finite wire and leads -
after the coupling at $t=0$, can be described by the Liouville-von
Neumann equation for the statistical operator $W(t)$. The evolution
of the finite wire itself can be captured by the reduced density 
operator $\rho (t)=\mathrm{Tr_LTr_R}W(t)$ (RDO), {\it i.\ e.}\ by
averaging over the the lead variables.  From the resulting integro-differential 
equation we retain only the lowest order (quadratic) terms in $H_T$ and
obtain \cite{NJP1,NJP2}
\begin{eqnarray}
      \nonumber
      {\dot\rho}(t)=&-&\frac{i}{\hbar}[H_\mathrm{S},\rho(t)]\\ 
      &-&\frac{1}{\hbar^2}\sum_{l=\mathrm{L,R}}\int dq\:\chi_l(t)
      ([{\cal T}_l,\Omega_{ql}(t)]+h.c.),
      \label{GME}
\end{eqnarray}
where we have introduced two operators to compactify the notation
\begin{eqnarray}
\nonumber
      &&\Omega_{ql}(t)=U_\mathrm{S}^\dagger (t) \int_{t_0}^tds\:\chi_l(s)
      \Pi_{ql}(s)e^{i((s-t)/\hbar )\varepsilon_l(q)}U_\mathrm{S}(t),\\\nonumber
      &&\Pi_{ql}(s)=U_\mathrm{S}(s)\left ({\cal T}_l^{\dagger}
      \rho(s)(1-f_l)-\rho(s){\cal T}_l^{\dagger}f_l\right )U_\mathrm{S}^\dagger(t),
\end{eqnarray}
with $U_\mathrm{S}(t)=e^{i(t/\hbar )H_\mathrm{S}}$,
and a scattering operator ${\cal T}$ acting in the many-electron Fock space of the system
\begin{eqnarray}
      \nonumber 
      {\cal T}_l(q)&=&\sum_{\alpha,\beta}{\cal T}_{\alpha\beta}^l(q)
      |{\bf \alpha}\rangle\langle {\bf \beta}| \ , \\
      {(\cal T}_l(q))_{\alpha\beta}&=&\sum_aT^l_{aq}\langle {\bf \alpha}
      |d_a^{\dagger}|{\bf \beta}\rangle \ .
\label{Toperator}
\end{eqnarray}
With the RDO it is possible to calculate the statistical average of the
charge operator $Q_\mathrm{S}=e\sum_a d_a^{\dagger}d_a$ for the coupled system
\begin{eqnarray}
      \nonumber
      \langle Q_\mathrm{S}(t)\rangle&=&{\rm Tr}\{W(t)Q_\mathrm{S}\}
      ={\rm Tr}_\mathrm{S}\{ [{\rm Tr}_\mathrm{LR}W(t)]Q_\mathrm{S} \}\\
      &=&{\rm Tr}_\mathrm{S}\{\rho(t)Q_\mathrm{S} \}
      =e\sum_{a,\mu} i^{\mu}_a \, \langle\mu | \rho(t) | \mu\rangle ,
\end{eqnarray}
with the traces assumed over the Fock space.
The average time-dependent spatial distribution of the charge can also be obtained,
\begin{equation}
      \langle Q_\mathrm{S}({\bf r},t)\rangle = e\sum_{ab}\sum_{\mu\nu}
      \Psi^*_a({\bf r})\Psi_b({\bf r})\rho_{\mu\nu}(t)\langle\nu |d^\dagger_ad_b|\mu\rangle .
\label{Qxy}
\end{equation}
The net current flowing into the sample is
\begin{eqnarray}
      \label{current}
      \langle J(t)\rangle&=&\langle J_\mathrm{L}(t)\rangle
      -\langle J_\mathrm{R}(t)\rangle\\ 
      \nonumber
      &=&\frac{d\langle Q_\mathrm{S}(t)\rangle}{dt}
      =e\sum_a \sum_{\mu} i^{\mu}_a \, \langle\mu | \dot\rho(t) | \mu\rangle \,.
\end{eqnarray}
The total current in Eq.\ (\ref{current}) is given by the left hand side of the 
GME, Eq.\ (\ref{GME}), and the partial currents associated to each SES and each lead
correspond to the terms of the sum on the right hand side (the trace of
the commutator of $\rho$ and $H_\mathrm{S}$ is zero).

The functions which modulate the coupling between the leads and the
quantum wire are built in the following way: For $0 \leq t < T_l$ we use
the Fermi-like function $f(t)=(e^{t\gamma}+1)^{-1}$, with $\gamma = 1.0$
ps$^{-1}$, and we define $\chi_l(t)=1-2f(t)$, where $l=\mathrm{L,R}$ is
the lead index.  Then, for $t \geq T_l$, $\chi_l(t)$ become step-like
functions alternating between 0 and 1, both with the same period $T$,
but with a delay of $T/2$.  In this way we mimic the on/off contact
switching done in the turnstile experiments.  We choose $T_\mathrm{L} <
T_\mathrm{R}=T_\mathrm{L}+T/2$, which means we first switch off the left contact
while the right one is still on. Then the left is turned on again while the
right is turned off, and so on.

\section{Numerical calculations, results, and discussion}
\label{calc}

\subsection{No magnetic field}
\label{Beq0}

\begin{figure}[htp]
\centering {\includegraphics[width=7.5cm]{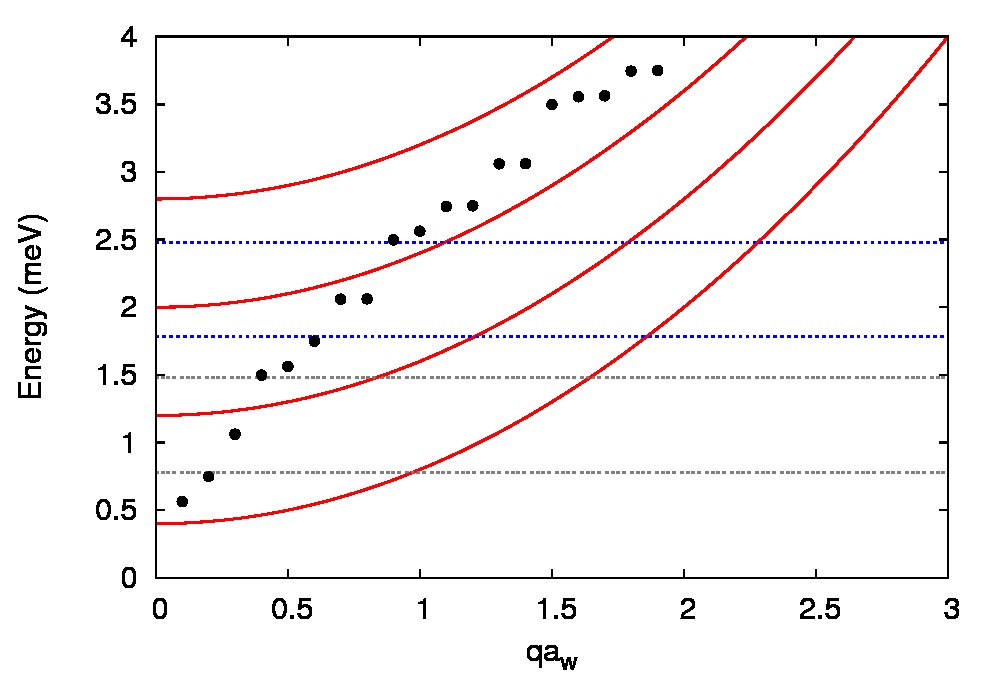}}
\caption{(Color online) The energy spectrum of the leads vs.\ the scaled wave vector $qa_w$.
        The subband index is $n=0,1,2,3$. With dots the energy spectrum of the
        isolated sample. With horizontal wide dashed lines the chemical potentials
        $\mu_L=1.48$ meV and $\mu_R=0.78$ (BW1), and with horizontal
        narrow dashed lines $\mu_L=2.48$ meV and $\mu_R=1.78$ meV (BW2).}
\label{Elev}
\end{figure}

The energy spectrum of the leads and of the sample are shown in Fig.\
\ref{Elev}.  The leads have parabolic subbands while the sample has
discrete levels.  The maximum energy for each subband shown in the graph
indicates the corresponding maximum wave vector in the $qa_w$-integration
of the GME.  The chemical potentials in the leads defining the bias
window (BW) are shown with the dashed horizontal lines.  We consider two BW's:
BW1 with chemical potentials $\mu_{L}=1.48$ meV and $\mu_{R}=0.78$
meV, and BW2 with  $\mu_{L}=2.48$ meV and $\mu_{R}=1.78$ meV respectively.
In both cases the applied bias is $eV_{\mathrm{bias}}=\mu_{L}-\mu_{R}=0.70$  meV.
In the numerical calculations of the reduced statistical operator we also
include the sample states with energy outside the BW, between
the limits $\mu_{R}-\Delta$ and $\mu_{L}+\Delta$ with $\Delta=0.1$ meV.
The first active window contains 4 SESs and the second one contains 5 SESs.

\begin{figure}[htp]
\centering
{\includegraphics[width=7.5cm]{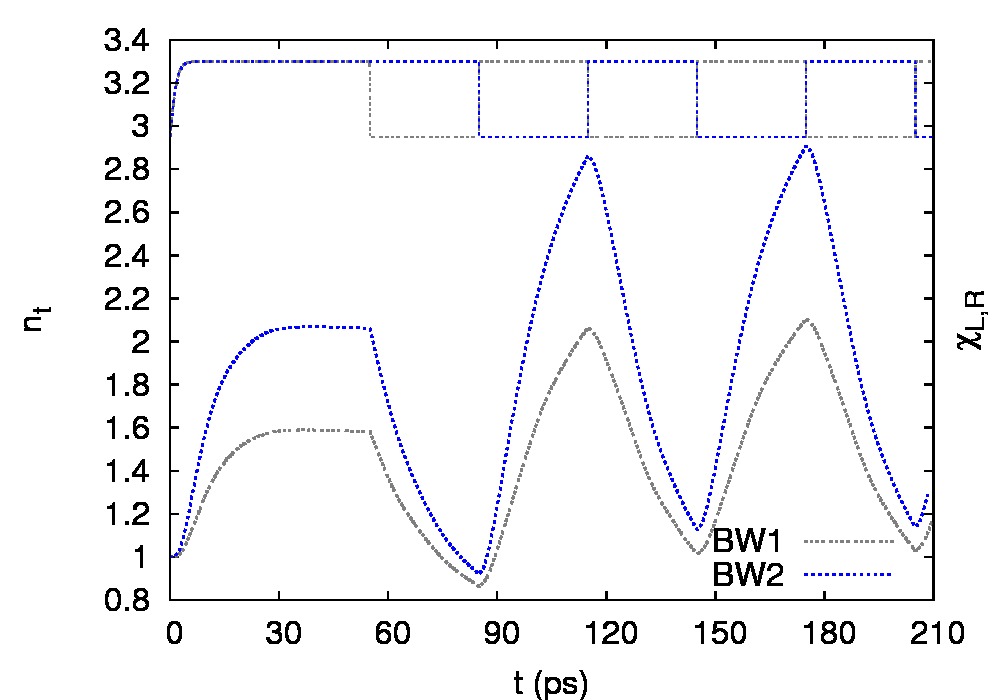}}
\caption{(Color online) The time dependent total occupation of the relevant SES for
        a system occupied initially with 1 electron ($\mu_0=2$) for the first (solid),
        and second (dotted) bias window. The time coupling functions $\chi_l(t)$, with
        $l=L,R$, are shown for reference. Other parameters: $g_0a_w^{3/2} = 926$ meV,
        $\delta_1a_w^2=1.0$, $\delta_2a_w^2=2.0$, and $T=60$ ps. In each case there is one
        electron on the lowest relevant state at the initial moment.}
\label{occup1}
\end{figure}

In Fig.\ \ref{occup1} we show the time-dependent total occupation $n_t$ of the
relevant (active) SES for BW1 and BW2.  The electrons occupying the
states situated below the active windows are not included, but only
those on the relevant states (or within the active windows).  In this
example the parameters characterizing the coupling of the sample
to the leads are $g_0a_w^{3/2}=926$ meV (with $a_w$ measured in
nm), $\delta_1a_w^2=1.0$, and $\delta_2a_w^2=2.0$.  The large value
for $g_0a_w^{3/2}$ is consistent with the small overlap of the wave
functions from the lead and from the sample mandated by the large values
for $\delta_1a_w^2$ and $\delta_2a_w^2$.  The largest contribution to
the overlap comes from the ``contact'' area within one $a_w$ around the
ends of the finite quantum wire at $x=\pm L_x$.  The unusual dimension
of $g_0$, which is $\mathrm{energy} \ \times \ \mathrm{length}^{-3/2}$
is a result of different dimensions of the wave functions in the lead and
in the sample: the former is $\mathrm{length}^{-1/2}$, being unbounded in
the direction along the lead, and the later is $\mathrm{length}^{-1}$.
We also select $\Delta_E=0.25\hbar\Omega_w$ in Eq.\ (\ref{gl}).
The time-dependent switching functions $\chi_l(t)$ are also shown in
the figure, with period $T=60$ ps, for $l=\mathrm{L,R}$.

Comparing the charge oscillation for the first and for the second active
window we see that more charge is transferred through the system for
BW2 than for BW1. There are two reasons for that.  One reason is that
BW2 includes three subbands of the leads, whereas BW1 includes only two
(Fig.\ \ref{Elev}).  The second reason is that the energy dispersion
in each subband increases with increasing energy, such that the
electrons have higher speed in BW2 than in BW1, and thus an increased
contribution to conduction.

\begin{figure}[htp]
\centering
{\includegraphics[width=7.5cm]{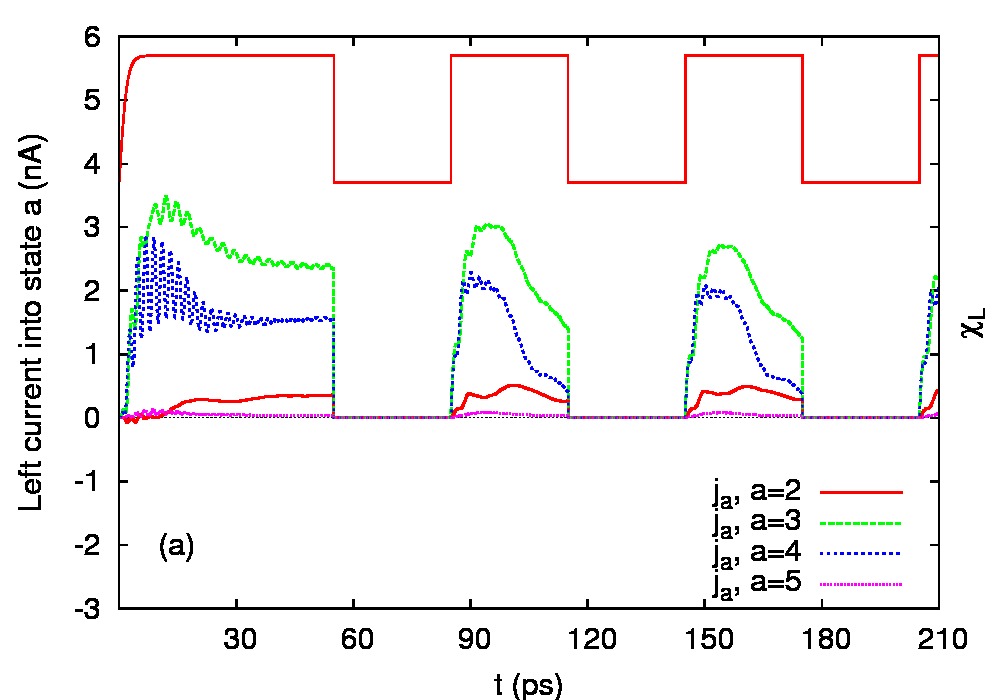}}
{\includegraphics[width=7.5cm]{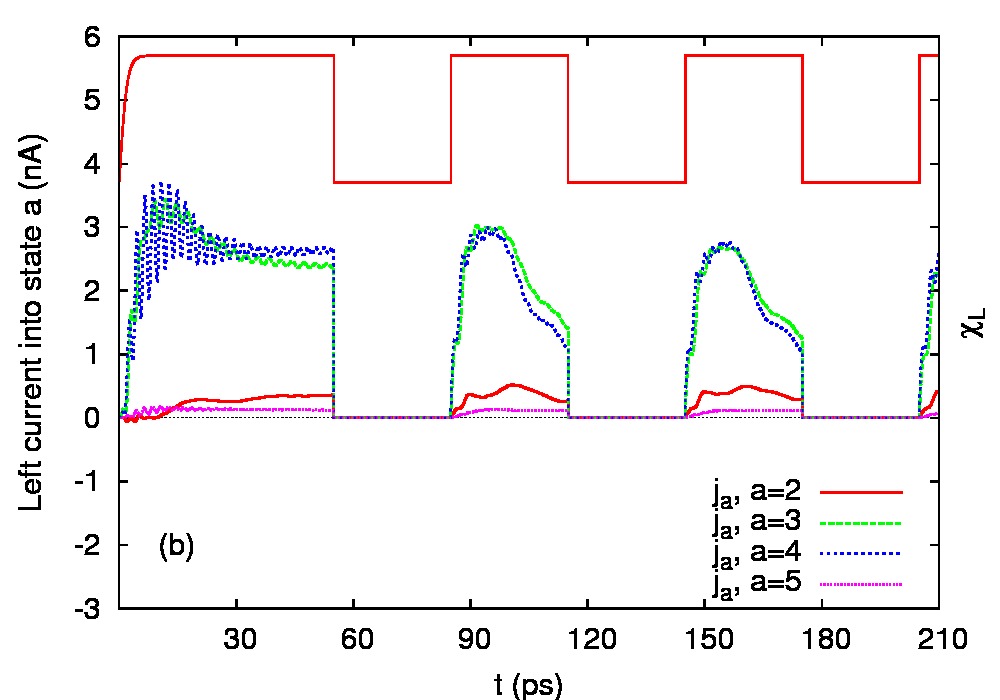}}
\caption{(Color online) The partial current entering the system from the left lead when
        the bias window contains one respectively two SES. Other parameters:
        $g_0 a_w^{3/2}=926$ meV, $\delta_1a_w^2=1.0$, $\delta_2a_w^2=2.0$, and
        $T=60$ ps.}
\label{current1}
\end{figure}

It is interesting to observe the behavior of the states situated at the
boundaries of the BW.  To show that we choose the BW1 and display in
Fig.\ \ref{current1}a the partial currents created by each sample state
included in the calculation.  These are the states 2,3,4,5 in the order
of the energy, shown with dots in Fig.\ \ref{Elev}.  State 1 (the lowest
dot) is considered totally occupied and frozen.  In this case state 2 is
slightly below the window, state 3 is within the window, and states 4 and
5 are slightly above the window.   In Fig.\ \ref{current1}b we show the
same partial currents, but now for a slightly higher chemical potential
of the left lead, $\mu_L=1.54$ meV, instead of $\mu_L=1.48$ meV used in
Fig.\ \ref{current1}a.  The state number 4 is now included in the BW
and consequently the corresponding current increases.  The current of
the state 5 also increases a little bit, while the current associated
to the other states does not change.  A similar behavior is displayed
by state number 9 situated on top of BW2 (not shown).

\begin{figure}[htp]
\centering
{\includegraphics[width=7.5cm]{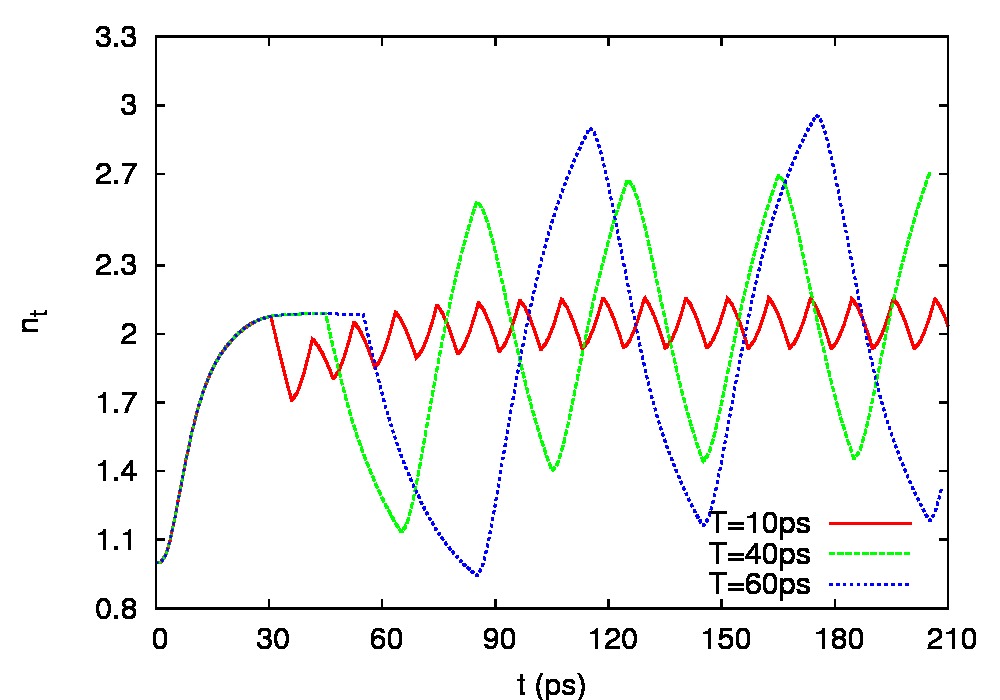}}
\caption{(Color online) The time dependent total occupation of the sample for
        three pulses at zero magnetic field. The trace for $T=60$ ps is the same
        as in Fig.\ \ref{occup1}. Other parameters: $g_0 a_w^{3/2}=926$ meV,
        $\delta_1a_w^2=1.0$, and $\delta_2a_w^2=2.0$.}
\label{occup2}
\end{figure}

The efficiency of the turnstile operation depends on the pulse length
$T$.  The previous results are obtained with $T=60$ ps. In those setups
the system transfers at least two electrons per cycle.  The present
GME method is valid in the lowest (quadratic) order of the lead-sample
coupling, which means the tunneling of the electrons from the leads to the
sample and back is a relatively slow process.  Therefore by increasing
or decreasing the pulse duration the transferred charge increases or
decreases respectively.  Denoting by $T_t$ the characteristic tunneling
time, if $T\leq T_t$ the turnstile operation is not expected to work,
the allowed time for the charging and discharging of the system
being too short.  This is the case for $T=10$ ps as shown in 
Fig.\ \ref{occup2}, when clearly very little charge can enter and leave
the system in a pumping cycle.  Actually, the time $T_t$ depends on
the pairwise coupling (or overlap) of each state of the leads to each
state of the sample with energies within the BW.  But in order to obtain
significant pumping effects, the pulse duration has to include the time
of flight (or propagation time) of electrons along the wire. This extra
time depends on the energy of the electrons injected from the left lead
all the way into the right lead.

Therefore, for a longer pulse, $T=40$ ps, the turnstile pumping process
is able to transfer charge through the sample.  The occupation number
has a triangular shape in time and it becomes periodic after 2-3 cycles.
During the initial cycle, which includes the initial charging phase, the
system accumulates more than 2 electrons and the steady state is already
reached at about 30 ps when the charge in the system is saturated.
This is possible because the right contact is still off.  The right
contact opens for the first time at 45 ps allowing more than 1 electron
charge to pass into the right lead.  For a longer period, like $T=60$
ps, the occupation number develops toward a saw-tooth profile typical for
the charging/relaxation processes.  The asymmetry of the charge peaks is
determined by the direction of the bias: the electrons leave the sample
faster than they entered. The system drives almost two electrons from one
lead into the other, which is remarkable given the length of our sample
(300 nm).

\begin{figure}[htp]
\centering
{\includegraphics[width=7.5cm]{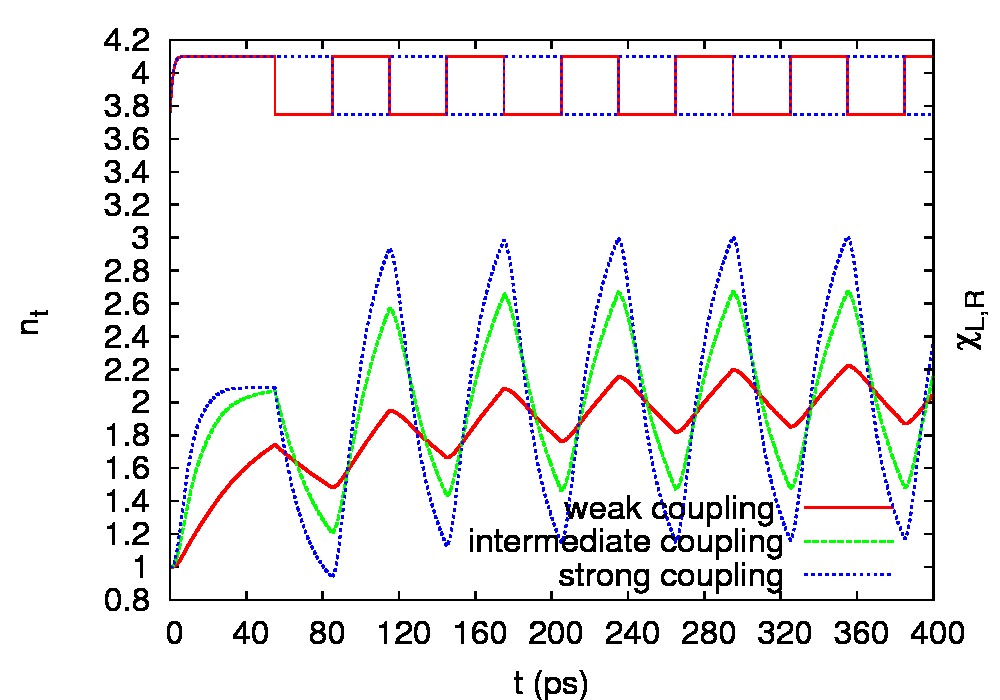}}
\caption{(Color online) The time dependent total occupation for $B=0$ T and two values of the coupling
        strength. Weak: $g_0 a_w^{3/2}= 1408$ meV, $\delta_1a_w^2=1.85$, $\delta_2a_w^2=3.7$.
        Intermediate: $g_0 a_w^{3/2}= 1408$ meV, $\delta_1a_w^2=1.39$, $\delta_2a_w^2=2.77$.
        Strong: $g_0 a_w^{3/2}= 1824$ meV, $\delta_1a_w^2=1.39$, $\delta_2a_w^2=2.77$.
        The pulse period is $T=60$ ps.}
\label{strengthB0}
\end{figure}
Another important aspect in our model is the strength of the lead-sample
coupling, {\it i.\ e.}\ the parameters $g_0$, $\delta_1$, and $\delta_2$ in Eq.\
(\ref{Toperator}).  Our GME implementation is restricted to the lowest
order in $H_T$ (quadratic), supplying the integro-differential equation (\ref{GME}),
and thus the parameters have to be appropriately selected.  In general it
is difficult to evaluate whether the coupling strength is sufficiently
low. A necessary (although not sufficient) condition is to obtain
positive diagonal elements of the statistical operator, which are the
populations of the MES and hence probabilities.  Although we always check in our
calculations, strictly speaking this condition does not guarantee 
the validity of the lowest order approximation (quadratic in $H_T$).  
So in practice
we cannot avoid choosing our parameters in a semi-empirical manner.
To have an idea about the relation between the pumping amplitude and the
coupling strength we show in Fig.\ \ref{strengthB0} three calculations
done with three strengths of the coupling, which we consider in relative
terms weak, intermediate, and strong.  In order to compare with the
results obtained in the presence of a magnetic field shown in the next
examples,  the scaled parameters $g_0a_w^{3/2}$ and $\delta_{1,2}a_w^2$
are chosen such that the {\em physical} values $g_0$ and $\delta_{1,2}$
are the same as for $B=0.9$ T.  The time dependent occupation of the states
in the active window is shown for longer times than in the previous figures 
to indicate better the final periodic regime.  It is not surprising
to see that the pumping amplitude increases with the coupling strength, since
tunneling of electrons becomes more likely.  The same can be said about the
current, which is essentially the time derivative of the charge, and it is
visible in Fig.\ \ref{strengthB0} that the slope of the charge increases
with the coupling strength.  The increase of the transient current with
the coupling strength has also been shown recently by Sasaoka {\it et al.}
although in a quantum dot pumping device. \cite{Sasaoka}

\subsection{Magnetic field present}
\label{Bne0}

Following our ansatz used to describe the system-leads coupling,
Eqs.\ (\ref{T_aq})-(\ref{gl}), we can visualize systems where
the physical parameters $g_0$, $\delta_1$, and $\delta_2$ should be
assumed constant, and others where the scaled values $g_0a_w^{3/2}$,
$\delta_1a_w^2$, and $\delta_2a_w^2$ should be kept constant with changing
values of the magnetic field rather.  In the following we explore both
possibilities. Our choice of $\Delta_E=0.25\hbar\Omega_w$ depends also
on the value of the magnetic field, but we have checked that variations
of this parameter between a scaled version and a fixed physical one
will only lead to vanishing quantitative changes in calculations where
we consider a range in the $q$ integration of the GME (\ref{GME}) that
includes 4 subbands of the leads. This choice of $\Delta_E$ guarantees
always the same number of subbands in the calculation.

\begin{figure}[htp]
\centering
{\includegraphics[width=7.5cm]{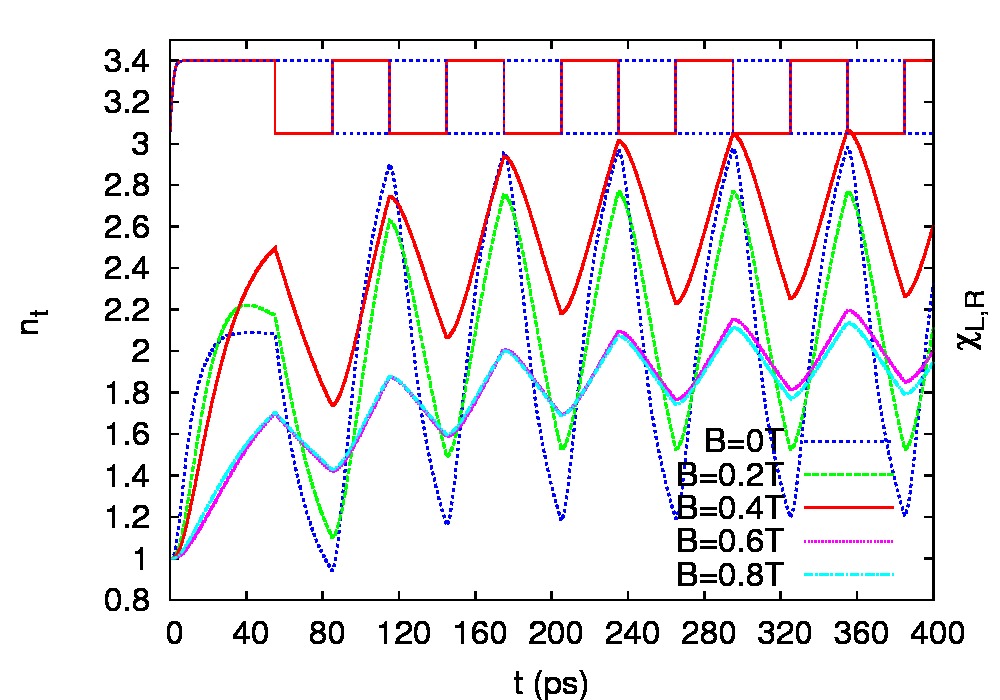}}
\caption{(Color online) The time dependent total occupation of the relevant SESs for  
        different values of the magnetic field. $g_0a_w^{3/2} = 926$ meV,
        $\delta_1a_w^2=1.0$, 
        $\delta_2a_w^2=2.0$, and $T=60$ ps.}
\label{occupB1}
\end{figure}

In the presence of a magnetic field, in a first approximation one might expect
the amplitude of the charge oscillations to decrease.  The reason being
that the electronic trajectories bend due to the Lorentz force and
the electrons might return to the source lead rather than traveling
directly to the drain lead. At the same time we know that a magnetic 
field generally reduces backscattering. In Fig.\ \ref{occupB1}
where we show the time dependent total charge for different values of
the magnetic field we see a reduction of the charge oscillations with
increasing magnetic field.  For comparison we include the $B=0$ case,
also shown in Fig.\ \ref{occup1} and Fig.\ \ref{occup2}.  To compare 
the results with and without magnetic field we initially keep the same
coupling parameters as measured with respect to the modified magnetic length,
$a_w$, i.\ e.\ $g_0a_w^{3/2}= 926$, meV $\delta_1a_w^2=1.0$, and
$\delta_2a_w^2=2.0$. The modified magnetic length $a_w$ decreases with
increasing magnetic field, as do the features of the wave functions.    
We select the bias windows like BW2, with three states included in 
the window and two marginal states in
the extended regions.  The energy levels and the wave functions depend on
the strength of the magnetic field both in the sample and in the leads.
In order to make the comparison shown in Fig.\ \ref{occupB1} meaningful
for each value of the magnetic field we shift the chemical potentials
in the leads such that the same energy levels of the quantum wire sample
are contained in the bias window with a fixed width.

The efficiency of the turnstile pump tends to decrease with increasing
the magnetic field.  For $B=0$ an amount of charge $Q_p \approx 1.8$
electron can be transmitted along the wire sample in one cycle, and
for $B=0.2$ T $Q_p \approx 1.2$.  For stronger magnetic fields $Q_{p}$
drops to 0.8 for $B=0.4$ T, and $Q_{p}<0.5$ at $B>0.6$ T.  It is also
clear that, at least for the present coupling parameters, the charging
time increases in the presence of the magnetic field.  During the initial
cycle, {\it i.\ e.}\ for $t<60$ ps, there is less charge accumulated in
the system in the normal switching regime than for $B=0$, and also the
charging process continues even after the pumping begins.  The periodic
regime is reached after a number of cycles which increases with $B$.

\begin{figure}[htp]
\centering
{\includegraphics[width=7.5cm]{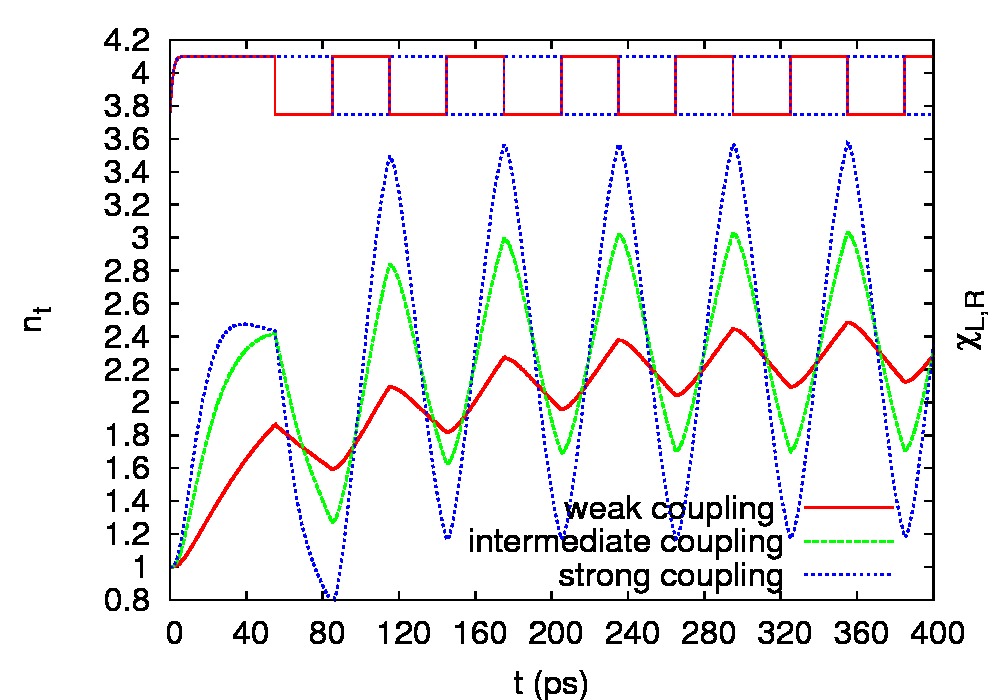}}
\caption{(Color online) The time dependent total occupation for $B=0.9T$ and three coupling
        strength: weak ($g_0a_w^{3/2} = 926$ meV, $\delta_1a_w^2=1.0$, and $\delta_2a_w^2=2.0$, 
        intermediate  ($g_0 a_w^{3/2}= 926$ meV, $\delta_1a_w^2=0.75$, and $\delta_2a_w^2=1.5$),
        and strong ($g_0 a_w^{3/2} = 1200$, $\delta_1a_w^2=0.75$, and $\delta_2a_w^2=1.5$).
        The pulse period is $T=60$ ps.}
\label{occupBcoupling}
\end{figure}

Although the presence of the magnetic field reduces the electron transfer,
the pumped charge still increases with increasing coupling between the
leads and the sample.  To show that we solve the GME for a fixed magnetic
field $B=0.9$ T for three sets of coupling parameters, which we again call
(in relative terms) weak, intermediate, and strong coupling, respectively,
see Fig.\ \ref{occupBcoupling}.  The parameters for the weak coupling are
the same as in Fig.\ \ref{occupB1}.  For the intermediate coupling we use
$g_0 a_w^{3/2}=926$ meV, $\delta_1a_w^2=0.75$, and $\delta_2a_w^2=1.5$.
For the strong coupling $g_0a_w^{3/2}=1200$ meV, $\delta_1a_2^2=0.75$,
and $\delta_2a_w^2=1.5$.  These parameters correspond to the same physical
parameters that were used in Fig.\ \ref{strengthB0}. So, if we now compare the results
at $B=0.9$ T with the results at $B=0$ we see quite similar charge
amplitudes, but somewhat more sensitive to the coupling strength at $B=0.9$ T.
For example, at $B=0.9$ T we obtain $Q_p \approx 0.4$ electrons at low
coupling, $Q_p \approx 1.2$ at intermediate coupling, and $Q_p \approx
2.4$  at strong coupling.  For $B=0$ these numbers are $Q_p \approx 0.3,
1.2$, and $1.8$, as can be read from Fig.\ \ref{strengthB0}.

\begin{figure}[htp]
\centering
{\includegraphics[width=7.5cm]{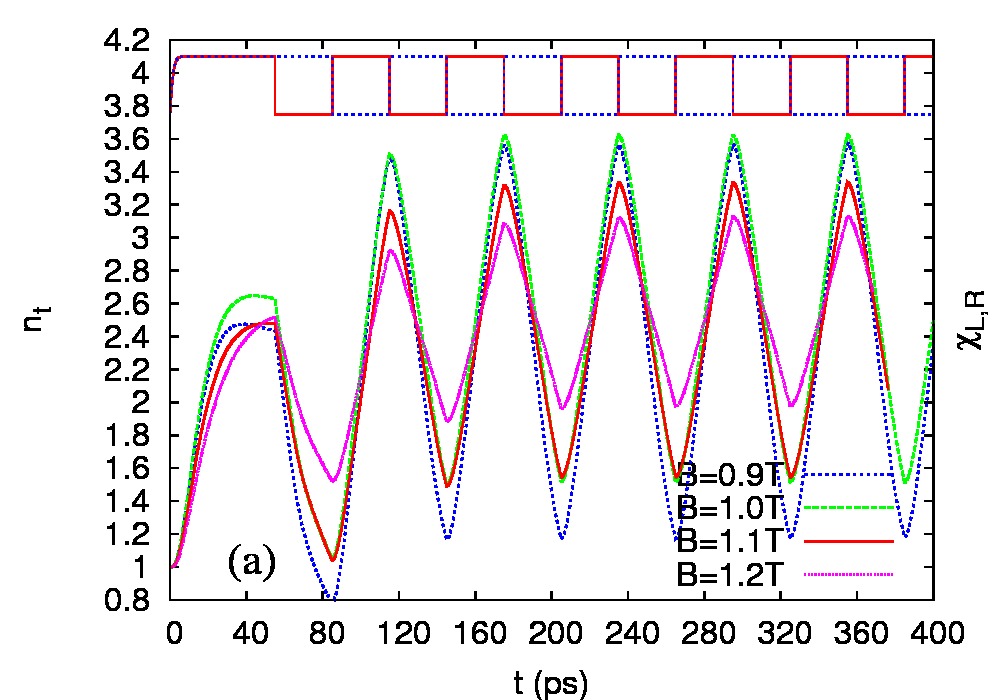}}
{\includegraphics[width=7.5cm]{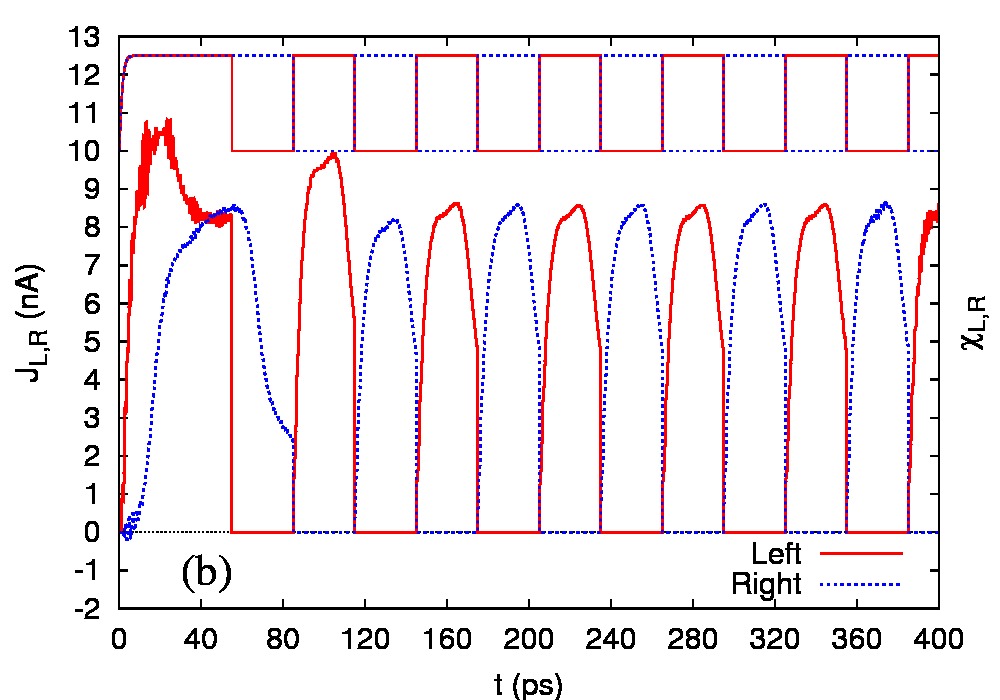}}
\caption{(Color online) (a) The time dependent total occupation of the relevant
        SESs  for $B=0.9-1.2$ T. (b) The total current entering the system from the left 
        lead and the total current exiting the system into the right lead for $B=1.0$ T.
        Parameters: $g_0 a_w^{3/2} = 1200$ meV, $\delta_1a_w^2=0.75$, and $\delta_2a_w^2=1.5$. 
        The pulse period $T=60$ ps.}
\label{occupBstrong}
\end{figure}

We believe the increased sensitivity of the pumping to the contact
strength in the presence of the magnetic field is a result of the edge
states created in the sample.  The edge states are indeed more and
more pronounced with increasing magnetic field, and so is the pumped
charge if the edge states are in good overlap with the coupling region.
Therefore with magnetic fields of about 1 T and strong coupling we can
obtain $Q_p$ of about 2 electrons, Fig.\ \ref{occupBstrong}, which is
slightly more than at zero magnetic field but with weak coupling $Q_p
\approx 1.75$, visible in Fig.\ \ref{occupBcoupling}.

The time dependent total currents in the left and right leads are shown in the lower
panel of Fig.\ \ref{occupBstrong} for $B=1$ T.  The currents suddenly
vanish at each contact when the contact is closed.  The currents have
a trapezoidal shape for the pulse duration $t=60$ ps, but they become
triangular for shorter pulses like $T=10$ ps (not shown).

\subsection{Analysis of edge states}
\label{ES}
 
\begin{figure}[htbq]
\begin{center}
      \includegraphics[width=0.23\textwidth,angle=0,viewport=10 22 350 250,clip]{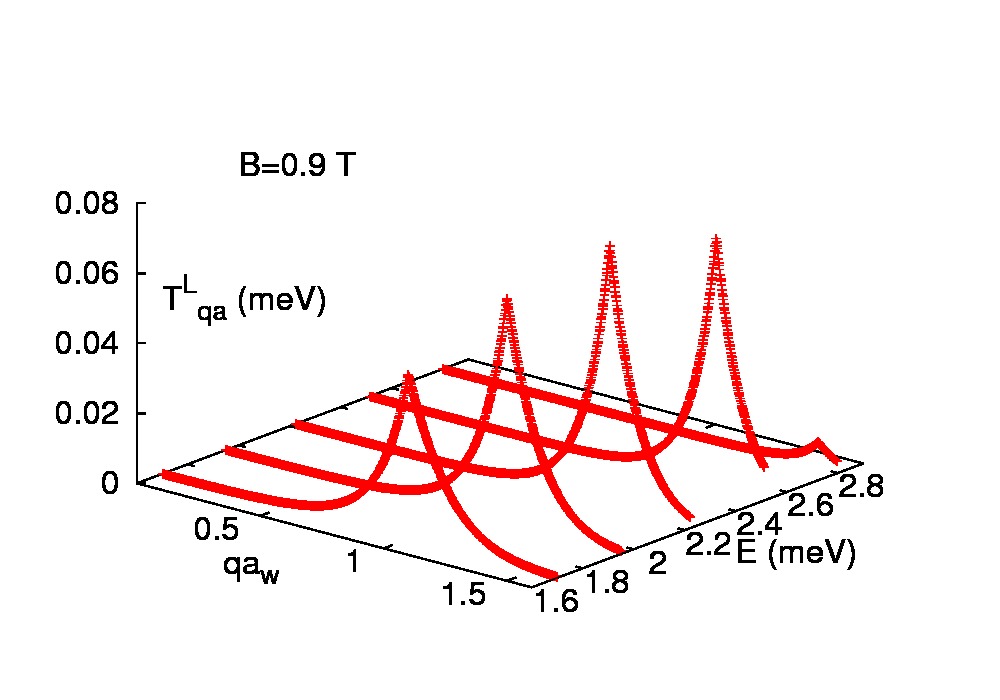}
      \includegraphics[width=0.23\textwidth,angle=0,viewport=10 20 350 250,clip]{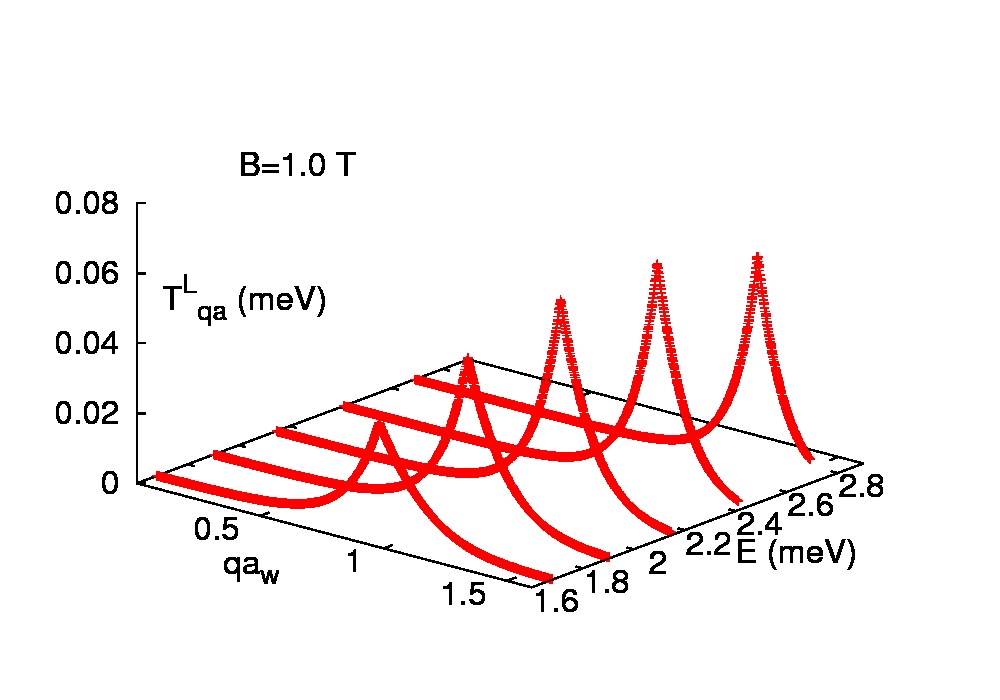}\\
      \includegraphics[width=0.23\textwidth,angle=0,viewport=10 20 350 250,clip]{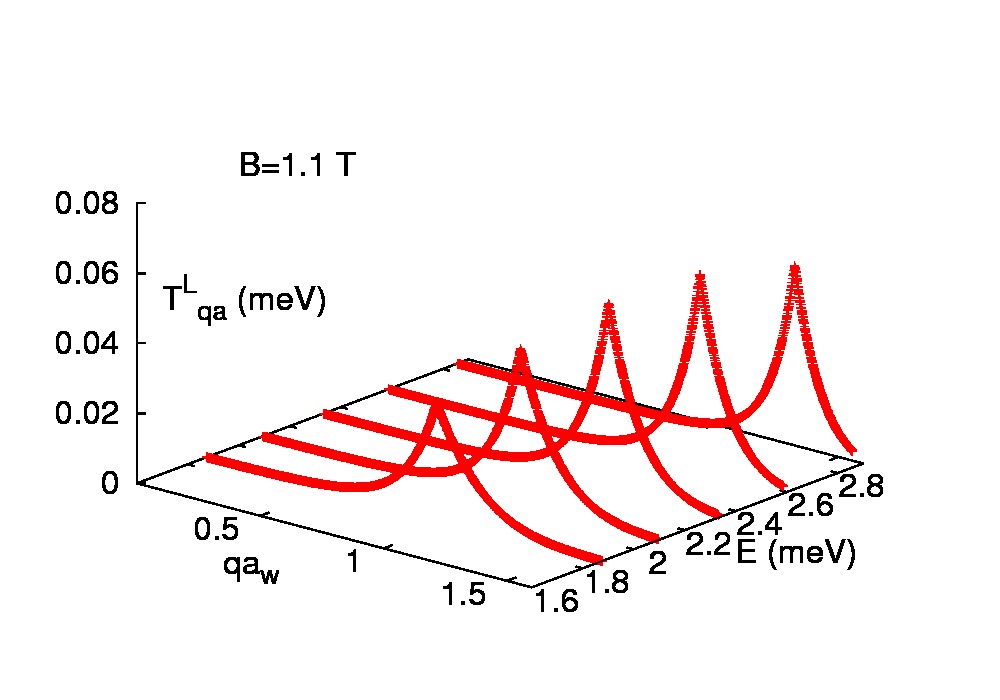}
      \includegraphics[width=0.23\textwidth,angle=0,viewport=10 20 350 250,clip]{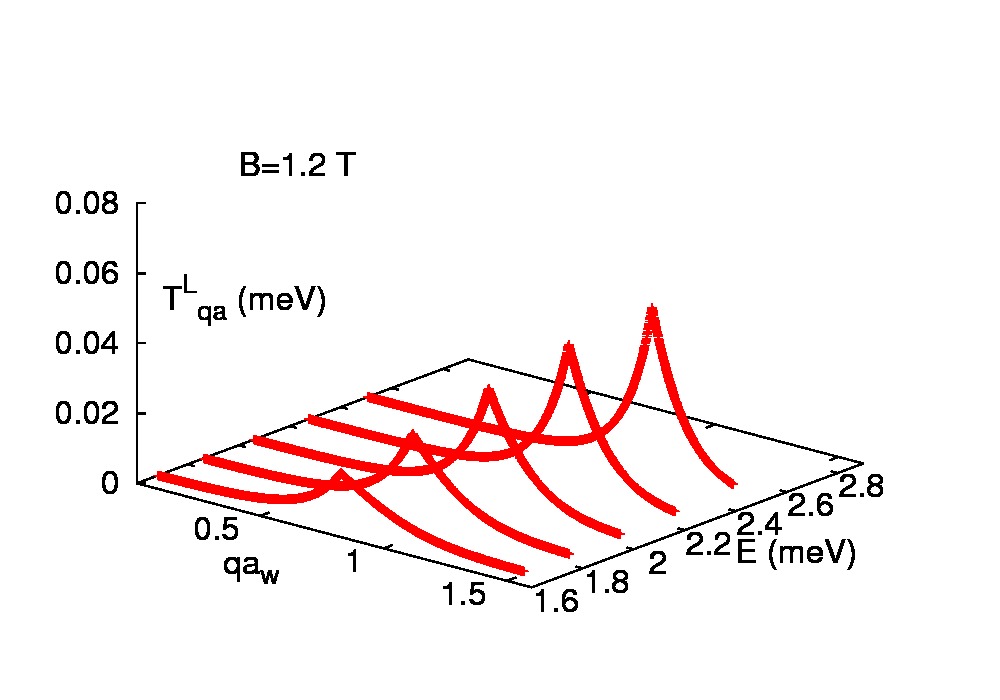}
\end{center}
\caption{(Color online) The coupling energies between the states $a=6-10$ and states of the 
        lowest subband in the 
        left (or the right) lead for $B=0.9 - 1.2$ T. 
        $g_0 a_w^{3/2} = 1200$ meV, $\delta_1a_w^2=0.75$, and 
        $\delta_2a_w^2=1.5$.} 
\label{T_L}
\end{figure}

Clearly, states in the sample with higher probability at the contact
edges contribute more to the pumping.   In Fig.\ \ref{T_L} we show the
coupling strength of each of the five states of the finite wire involved
in the transport for each magnetic field of Fig.\ \ref{occupBstrong},
only for the lowest subband of the leads.   The general trend of the
coupling coefficients is to decrease with increasing magnetic field.
So the decrease of the pumping when the magnetic field increases can be
explained by the decrease of the coupling strength.  The contribution
of each state to the transport is given by an integration over all the
lead states $q$ in the GME.

\begin{figure}[htp]
\centering
{\includegraphics[width=7.5cm]{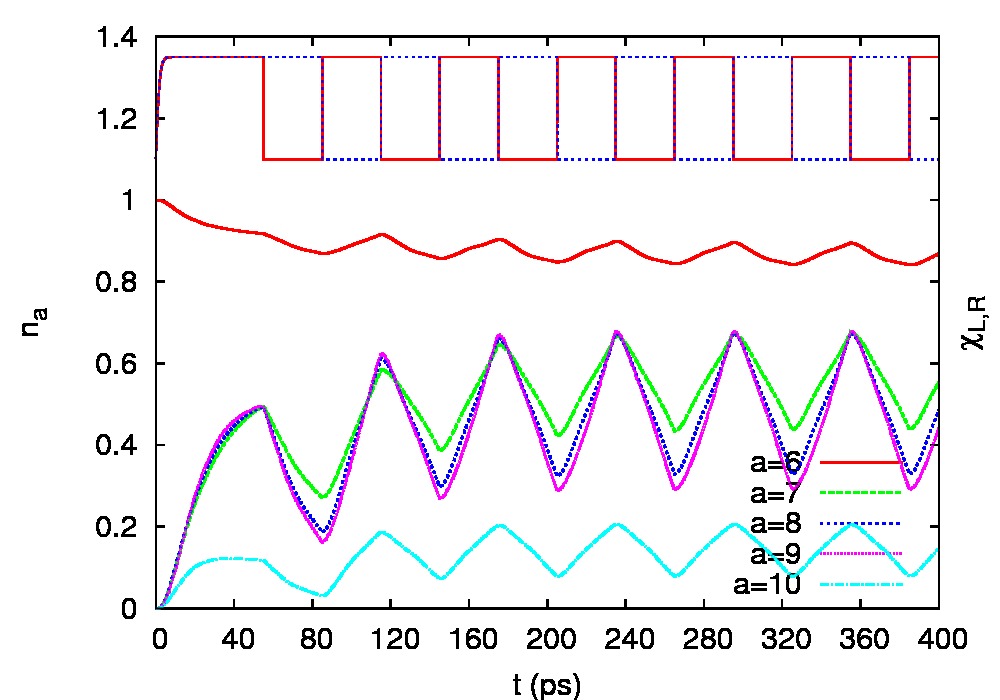}}
\caption{(Color online) The time dependent total occupation of the relevant SESs for $B=1.2$ T. 
        $g_0 a_w^{3/2} = 1200$ meV, $\delta_1a_w^2=0.75$, and $\delta_2a_w^2=1.5$. $T=60$ ps.}
\label{occupBpartial}
\end{figure}

To see the contribution of each state to the transport we show in Fig.\
\ref{occupBpartial} the time dependent occupation of all 5 states included
in the calculation for $B=1.2$ T, which are the states number 6-10 in the
single-electron energy spectrum of the finite wire.  It is clear that
the three middle states $a=7,8,9$ contribute unequally to the pumping.
These states are well inside the BW, but the coupling energies are
slightly different as seen in Fig.\ \ref{T_L} for $B=1.2$ T.  For the
other values of the magnetic field shown in Fig.\ \ref{T_L} the states
within the BW have nearly equal coupling to the leads and consequently
nearly equal contributions to transport.  So in general the coupling strength
may depend on the state and so does the corresponding partial current.
The states 6 and 10 included in Fig.\ \ref{occupBpartial}  are slightly
outside the BW and obviously their contribution to transport is smaller.

\begin{figure}[htp]
\begin{center}
\includegraphics[width=3.5cm,angle=0,viewport=20 10 250 180,clip]{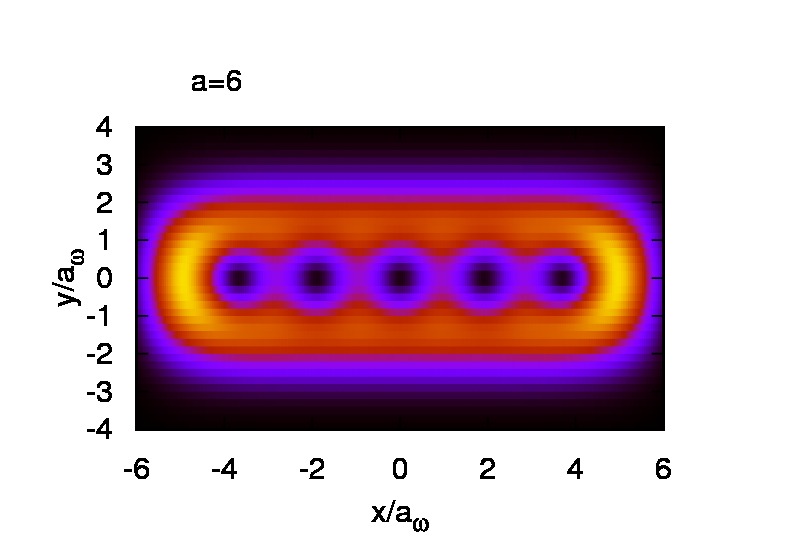}
\includegraphics[width=3.5cm,angle=0,viewport=20 10 250 180,clip]{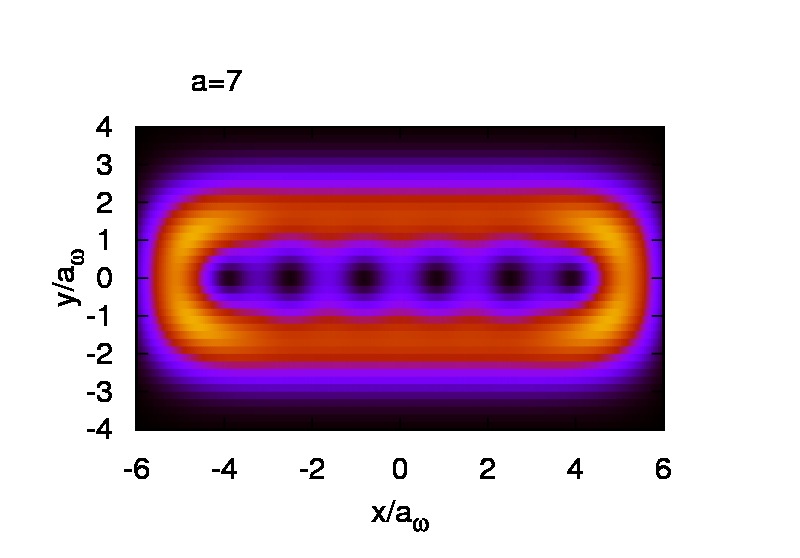}\\
\includegraphics[width=3.5cm,angle=0,viewport=20 10 250 180,clip]{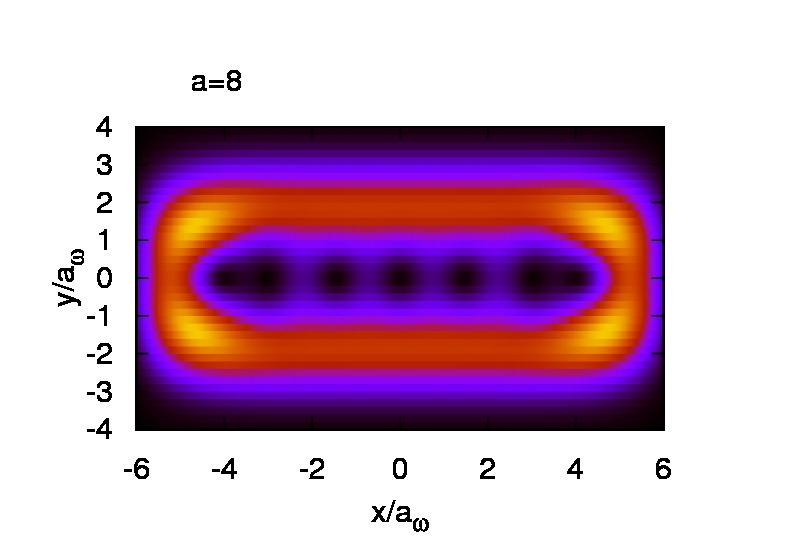}
\includegraphics[width=3.5cm,angle=0,viewport=20 10 250 180,clip]{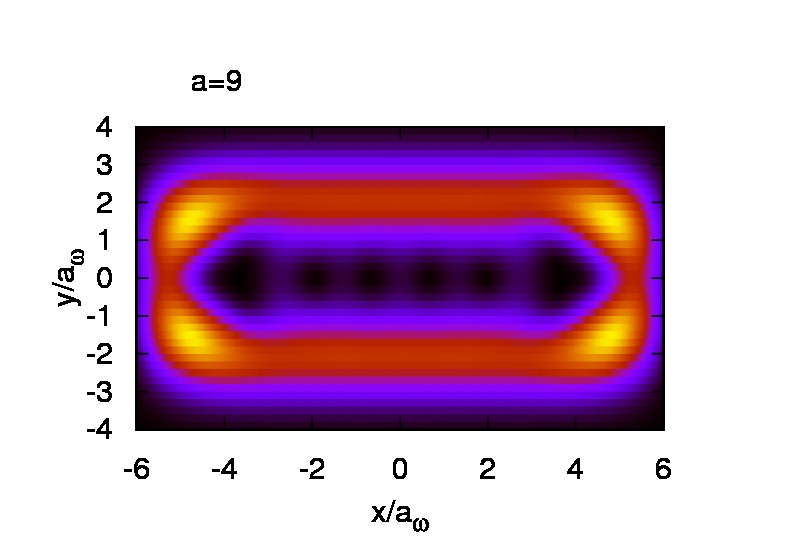}\\
\includegraphics[width=3.5cm,angle=0,viewport=20 10 250 180,clip]{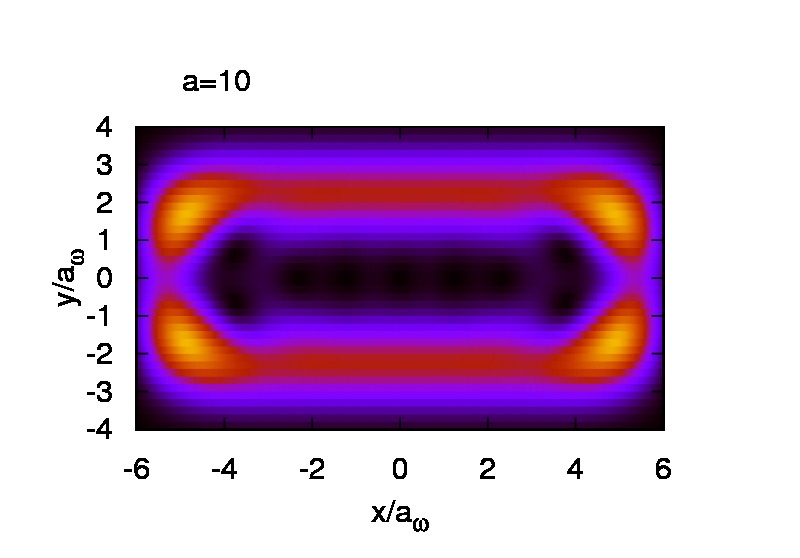}
\end{center}
\caption{(Color online) The probability density of the single-electron eigenstates of the sample  
        number 6-10 (top left - right - down), for $B=1$ T. }
\label{probdens}
\end{figure}

In Fig.\ \ref{probdens} we show the probability density associated to
the five active single-electron states (number 6-10) of the finite
quantum wire.  The figures indicates that all the five states have
the characteristics of an edge state. In vanishing magnetic field the
probability density of the active states is on the average smeared
over the finite wire. As the magnetic field increases the Lorentz force
squeezes the probability of some states close to the edge of the finite
wire. This also happens at the hard-wall ends of the wire, the contact
area. This fact explains why the increasing of the coupling through
increasing $g_0$ should be more effective at high magnetic field. From
Fig.\ \ref{probdens} it is evident that the edge states will have
different coupling strengths to the leads due to the difference in their
finer structure in the contact area. This finer structure in the contact
area of the wire induces differential coupling to the states in the
different subbands of the leads.

\begin{figure}[htp]
\includegraphics[width=0.15\textwidth,angle=0,viewport=20 45 230 210,clip]{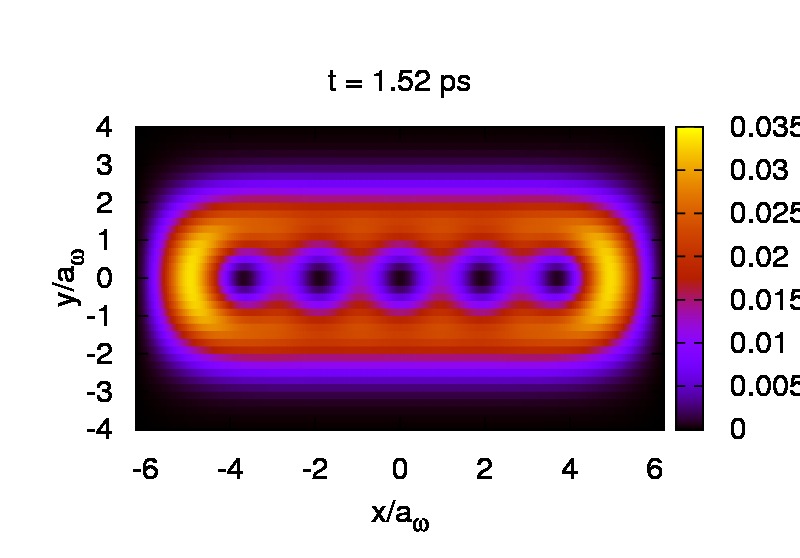}
\includegraphics[width=0.15\textwidth,angle=0,viewport=20 45 230 210,clip]{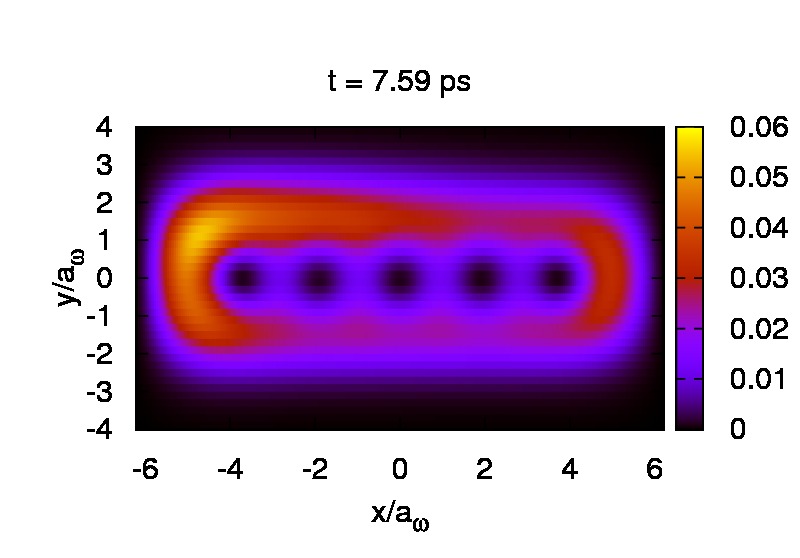}
\includegraphics[width=0.15\textwidth,angle=0,viewport=20 45 230 210,clip]{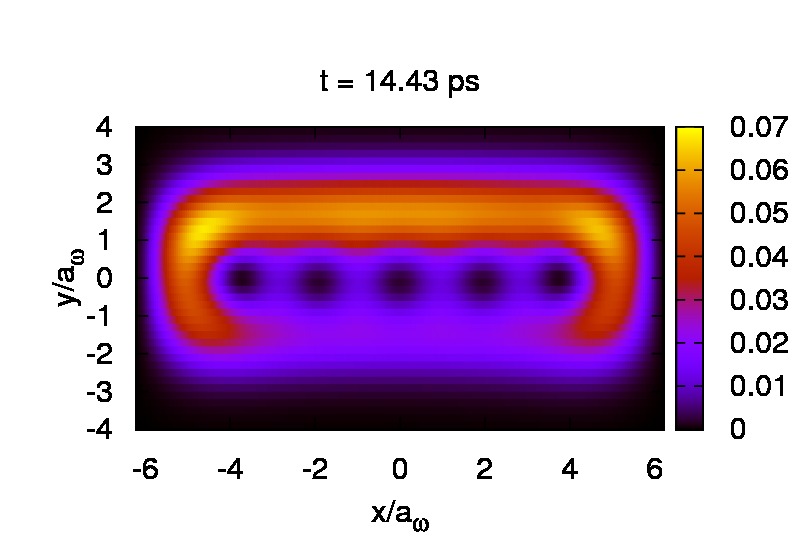}\\
\includegraphics[width=0.15\textwidth,angle=0,viewport=20 45 230 210,clip]{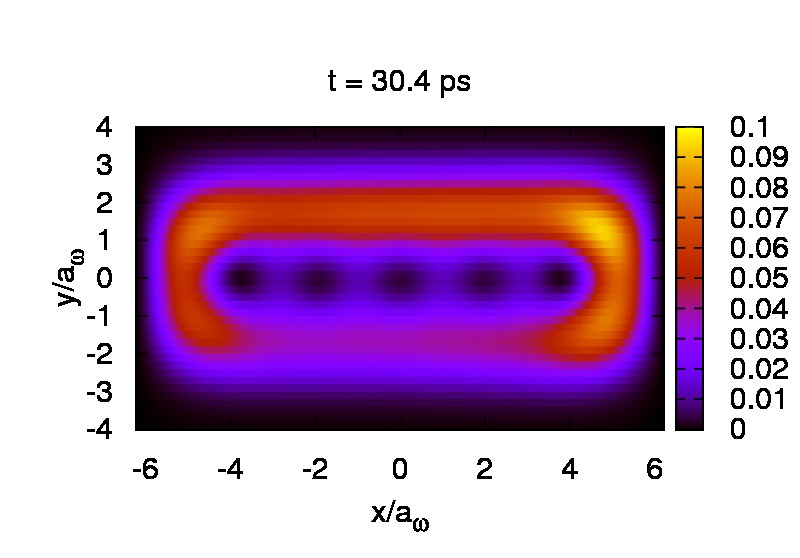}
\includegraphics[width=0.15\textwidth,angle=0,viewport=20 45 230 210,clip]{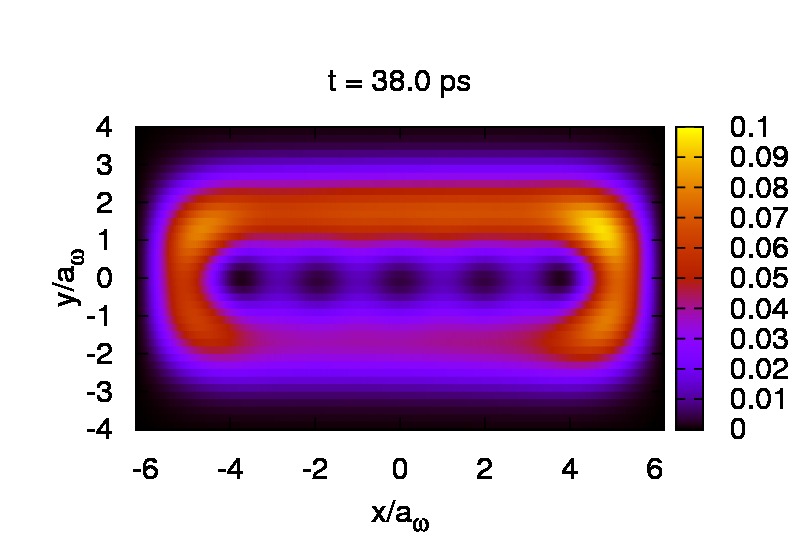}
\includegraphics[width=0.15\textwidth,angle=0,viewport=20 45 230 210,clip]{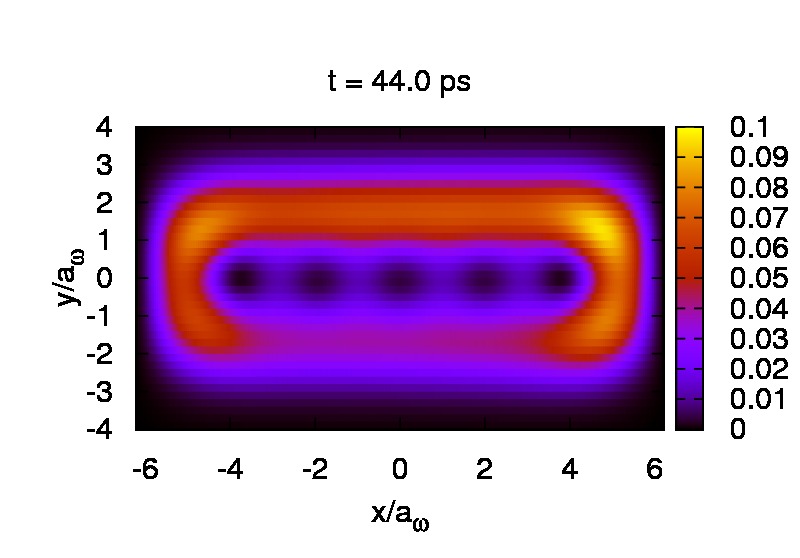}\\
\includegraphics[width=0.15\textwidth,angle=0,viewport=20 45 230 210,clip]{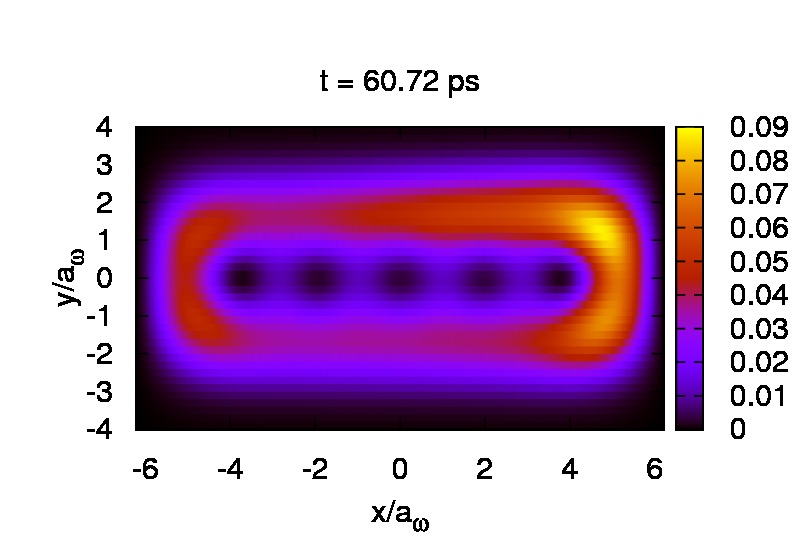}
\includegraphics[width=0.15\textwidth,angle=0,viewport=20 45 230 210,clip]{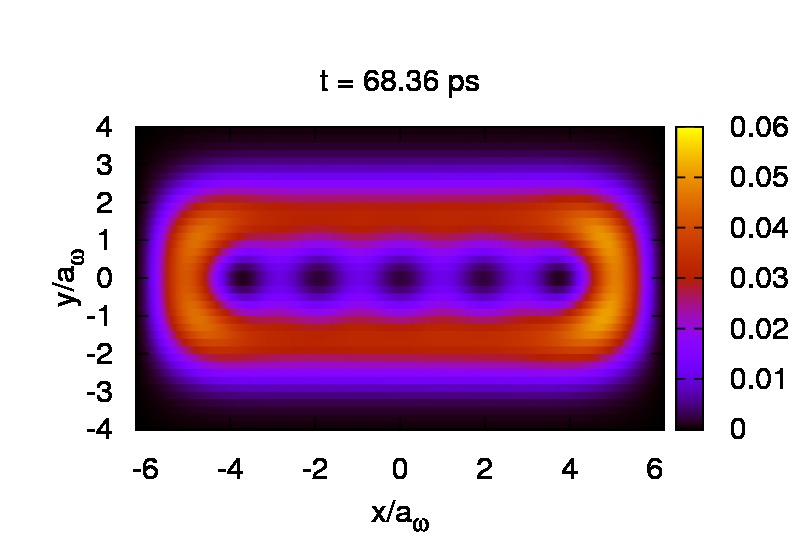}
\includegraphics[width=0.15\textwidth,angle=0,viewport=20 45 230 210,clip]{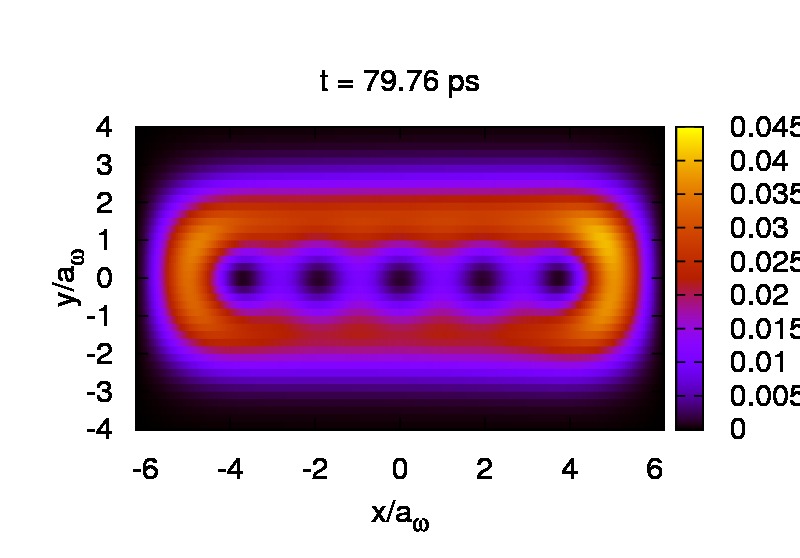}\\
\includegraphics[width=0.15\textwidth,angle=0,viewport=20 12 230 210,clip]{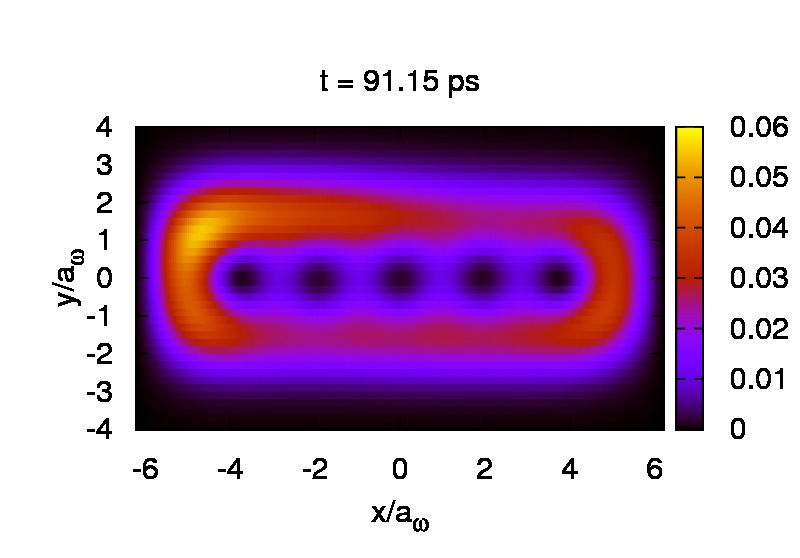}
\includegraphics[width=0.15\textwidth,angle=0,viewport=20 12 230 210,clip]{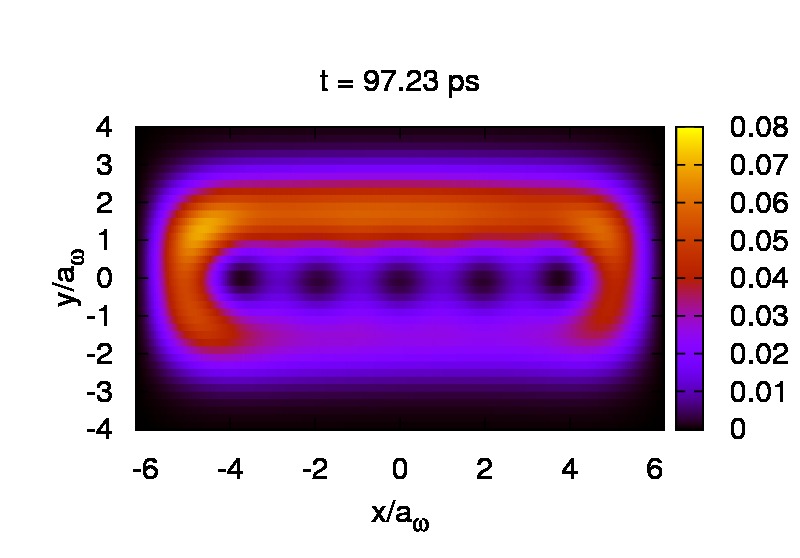}
\includegraphics[width=0.15\textwidth,angle=0,viewport=20 12 230 210,clip]{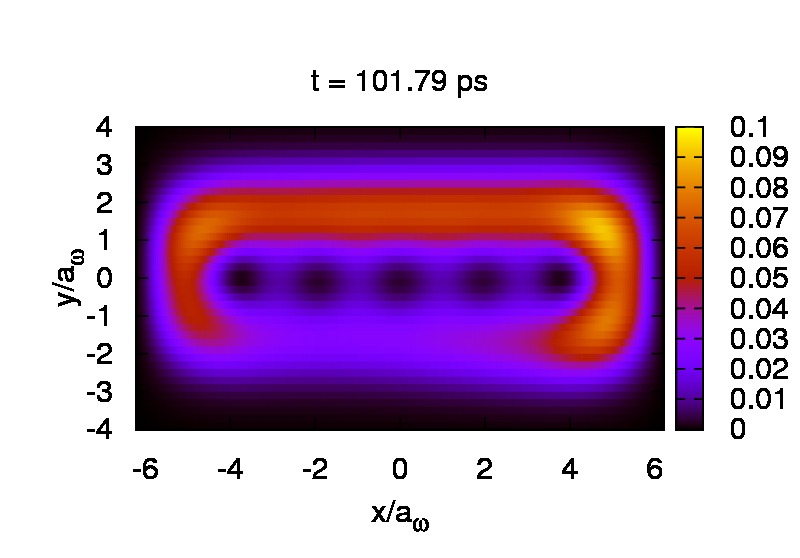}
\caption{(Color online) The average spatial charge distribution in the quantum wire for $B=1$ T with 
        strong coupling, at different moments of time.}
\label{chargedens}
\end{figure}

The time-dependent charge in the quantum wire is shown in Fig.\
\ref{chargedens} for the parameters used in Fig.\ \ref{occupBstrong}b.
The charge distribution reflects the geometry of the quantum wire
and of the five SESs involved, and indicates the propagation of the electrons in
the system.  The selected time moments cover the initial charging cycle
plus a part of the next cycle.  It is interesting to observe how the
electrons are injected at the left contact into the sample traveling
along the quantum wire on the upper edge channel, and how they are
reflected back at the right contact traveling along the lower channel.

\section{Conclusions}
\label{conc}

The turnstile conduction of a two-dimensional parabolic wire, seen as ``the
sample'', attached to  semi-infinite leads of a similar parabolic shape
has been studied using a non-Markovian generalized master equation method.
This system is far more complex than the simple two- or three-level system
with one-dimensional leads considered in an earlier publication based
on the Keldysh-Green functions approach.\cite{TSPPRB} The eigenstates
of the leads and of the finite wire have been calculated in an external
perpendicular magnetic field using a combination of analytical and
numerical methods for large functional basis sets.  We have taken into
account the subband structure of the leads to which it is attached. We
have also described phenomenologically the coupling coupling between
the states in the leads and the states in the finite wire as a nonlocal
overlap of the wave functions from both sides of the contact.

We have analyzed the effects of the bias window, pulse length, and
magnetic field on the evolution in time of the number of electrons in
the sample.  We have found that longer pulses are more favorable for
the turnstile pumping as the electrons need time to propagate along the
sample wire.  One or two electrons could be transferred through the 300
nm wire using pulses of 40 or 60 ps.

The comparison of the results obtained with and without a magnetic field
is a difficult issue.  If a magnetic field is present all energies shift,
all wave functions change (both in the sample and in the leads), and
also all elements of the coupling tensor $T_{qa}^l$ between lead and
sample states change.  Of course these changes depend on the strength
of the field.  Therefore it is difficult to create similar conditions
for two different field values, {\it i.\ e.}\ the same number of states
in the bias window and the same coupling energies, in order to compare
only the amplitude of the pumped charge.  To do that we have selected
the parameters describing the phenomenological coupling of the leads
to the finite wire in two different ways, both scaled and not scaled
with the effective width of the sample, which implicitly depends on
the magnetic field.  This is an issue that can only be better resolved
with a more involved microscopic model of the coupling and comparison
to experiments where the coupling strength could be varied, for example
by using finger-strip gates.

The charge distribution inside the system, Fig.\ \ref{chargedens},
emphasizes the dynamics induced by the charging/discharging sequences.
The charge propagation along edge-states in a strong magnetic field
indicates that the optimal turnstile period depends on the external
magnetic field.  Experimental studies of turnstile pumping in quantum
wires have to clarify the relationship between the magnetic field,
pumping amplitude, and contact strength.

\begin{acknowledgments}
The authors acknowledge financial support from the Research Fund of the
University of Iceland, the Development Fund of the Reykjavik University
(grant T09001), and the Icelandic Research Fund (Rannis).  V.M. also
acknowledges the hospitality of the Science Institute - University of
Iceland and the partial financial support from the Romanian Ministry of
Education and Research, PNCDI2 program (Grant No. 515/2009) and Grant No.
45N/2009.
\end{acknowledgments}

\end{document}